\documentstyle[12pt,aasms4]{article}
\begin{document}
\newcommand\Msun {M_{\odot}\ }
\newcommand\Lsun {L_{\odot}\ }

\title{
WFPC2 OBSERVATIONS OF LEO~A:\\ 
A PREDOMINANTLY YOUNG GALAXY WITHIN THE LOCAL GROUP\altaffilmark{1}}
\vskip0.5cm
\author{{\bf Eline Tolstoy}}
\affil{Space Telescope - European Coordinating Facility, 
Karl-Schwarzschild str. 2, 
D-85748 Garching bei M\"{u}nchen, 
Germany.}
\author{{\bf J.S. Gallagher, A.A. Cole, J.G. Hoessel}}
\affil{University of Wisconsin, Dept. of Astronomy, 475 N. Charter St,
Madison, WI 53706}
\author{{\bf A. Saha}}
\affil{NOAO, 950 N. Cherry Ave, Tucson, AZ 85726-6732}
\author{{\bf R.C. Dohm-Palmer, E.D. Skillman}}
\affil{University of Minnesota, Dept. of Astronomy, Minneapolis, MN 55455}
\author{{\bf Mario Mateo, D. Hurley-Keller}}
\affil{University of Michigan, Dept. of Astronomy, 821 Dennison Building,
Ann Arbor, MI 48109-1090}
\slugcomment{Accepted for publication in Astron. J. (Sept 1998)}
\altaffiltext{1}{Based on observations with the NASA/ESA 
{\it Hubble Space Telescope} obtained at the Space Telescope
Science Institute, which is operated by AURA, Inc. under
NASA contract NAS 5-26555}

\begin{abstract}

The unprecedented detail of the WFPC2 colour-magnitude diagrams of the 
resolved stellar population of Leo~A presented here allows us to determine 
a new distance and an accurate star formation history for this extremely 
metal-poor Local Group dwarf irregular galaxy.  From the position of the 
red clump, the helium-burning blue loops and the tip of the red giant 
branch, we obtain a distance modulus, m$-$M=24.2$\pm$0.2, or 690 $\pm$ 60 
kpc, which places Leo~A firmly within the Local Group.  Our interpretation 
of these features in the WFPC2 CMDs at this new distance based upon 
extremely low metallicity (Z=0.0004) theoretical stellar evolution models 
suggests that this galaxy is predominantly young, {\it i.e.} $<$~2~Gyr old.  
A major episode of star formation 900$-$1500~Gyr ago can explain the red 
clump luminosity and also fits in with our interpretation of the number of 
anomalous Cepheid variable stars seen in this galaxy.  We cannot rule out 
the presence of an older, underlying globular cluster age stellar 
population with these data.  However, using the currently available stellar 
evolution models, it would appear that such an older population is limited 
to no more than 10\% of the total star formation to have occured in this 
galaxy.  Leo~A provides a nearby laboratory for studying young metal poor 
stars and investigations of metal-poor galaxy evolution, such as is 
supposed to occur for larger systems at intermediate and high redshifts.
\end{abstract}

\keywords{galaxies: irregular, Local Group, color-magnitude diagrams,
stellar content, distance }

\section{Introduction}

Small galaxies are common and apparently structurally simple.  They may 
provide an important perspective on how luminous structures have evolved in 
the Universe.  Despite a variety of theoretical models (e.g., Dekel \& Silk 
1986; Hensler \& Burkert 1990), we cannot predict how even the simplest 
galaxies have changed over time.  In particular the internal clocks which 
set the time interval for major star formation are seen to be highly 
variable in the Galactic retinue of dSph, ranging from ancient systems 
where star formation was complete 10 Gyr ago to galaxies that are mainly 
intermediate age (e.g., van den Bergh 1994).  The presence of numerous 
small, actively star forming field galaxies at moderate redshifts of 
0.3$<$z$<$1 suggests that such asynchronous behavior may be the rule rather 
than the exception amongst smaller galaxies (e.g., Babul \& Ferguson 1996).

However, it still remains to be understood exactly what are faint blue 
galaxies (FBGs) that are found in deep imaging surveys (e.g., Ellis 1997).  
Their sheer numbers make them a cosmologically significant population and 
an important tracer of the star formation history (SFH) of the universe.  
There is considerable evidence that these FBGs are predominantly 
intermediate redshift ($z<1$, or a look-back time out to roughly half a 
Hubble time), late type, intrinsically {\it small} galaxies, undergoing 
strong bursts of star formation (e.g., Glazebrook {\it et al.} 1996; Lilly 
{\it et al.} 1996; Odewahn {\it et al.} 1996).  The best way to understand 
FBGs is to discover their nearby counterparts which we can study in detail, 
and directly observe the absence or presence of past bursts in 
colour-magnitude diagrams (CMD).  Irregular galaxies are very strong 
candidates for what is left of the FBGs in our nearby universe.  They are 
systems which may have undergone one or more bursts of star formation in 
the last few Gyr (Matteucci \& Tosi 1985).  They exist in large enough 
numbers that, could they be made bright enough for a short time in the 
past, they could easily account for the population of FBGs (e.g., Babul \& 
Ferguson 1996).

It is still an open question - why dwarf galaxies in the Local Group 
display such a wide range in stellar age distributions.  In our sample 
(Skillman 1998) we are finding enhanced star formation rates (SFR) over Gyr 
time scales, during which the bulk of the stellar populations within small 
galaxies are formed, rather than very short discrete bursts ($\lesssim$ a 
few $\times 10^7$yrs).  The physical mechanisms responsible for this 
behaviour are uncertain.  Current theoretical models mainly consider cases 
where SFR is tied to the gas supply.  Epochs of active star formation can 
then occur either due to delayed cooling of gas associated with a small 
galaxy (e.g., Kepner, Babul, \& Spergel 1997, Spaans \& Norman 1997), or 
through rejuvenation of a pre-existing small galaxy due to gas capture 
within a galaxy group (e.g., Silk, Wyse, \& Shields 1987).  Another 
possibility is that dwarfs were formed from tidal debris produced by 
interactions between galaxies in the Local Group, although it is not clear 
what objects could have spawned the Leo I and considerably more distant 
Leo~A dwarfs (e.g., Hunsburger, Charlton \& Zaritsky 1996).

Here we present the results for Leo~A ($\equiv$ DDO~69, Leo III, 
UGC~5364), a nearby Magellanic dwarf irregular galaxy.  A variety of 
studies have given this galaxy a large range of possible distances (e.g., 
Hoessel {\it et al.} 1994 [hereafter, H94], and references therein).  There 
are several faint HII regions distributed along the ridge of highest column 
density HI (Tolstoy~1996; Hunter, Hawley \& Gallagher 1993), which provide 
a very low limit to the current SFR (over the last 10Myr) of $< 10^{-4} 
\Msun yr^{-1}$.  The brightest H$\alpha$ emission in the galaxy comes from 
a planetary nebula, which yields an extremely low oxygen abundance of 
$\sim$~2.4\% solar (Skillman, Kennicutt \& Hodge 1989).

Recent HI observations (Young \& Lo 1996), show that the optical galaxy is 
surrounded by a large HI halo, extending out $\sim$3 times the optical 
diameter at a column density of 4$\times 10^{19} \rm cm^{-2}$.  The 
detected HI flux corresponds to an HI mass of $(8.1\pm 1.5) \times 10^7 \rm 
~\Msun$, of which $\sim$30\% is in the halo at column densities below 
$2\times 10^{20} \rm cm^{-2}$.  The observed HI velocity gradient across 
Leo~A in HI is so small that the changes in the velocity dispersion most 
likely reflect the conditions in the ISM rather than rotation or velocity 
crowding effects.  These HI observations show that the physical state of 
the ISM in Leo A is surprisingly similar to that in other, larger, more 
metal-rich galaxies (including our own), despite the fact that Leo~A is 
dominated by internal motions with very little effect coming from the 
exceedingly low global rotation measures.

The resolved stellar population of Leo~A has been studied before by 
Tolstoy~(1996) in the Thuan-Gunn filter system.  It was concluded that the 
SFR in the galaxy must have been higher in past than at the present time by 
a factor of $\sim$~3.  Although the interpretation only went back 
$\sim$1~Gyr because of the limitations of crowding and sensitivity of the 
ground based images.  This study adopted the distance as determined by H94.  
Our present observations favor a much smaller distance (see \S 3.1), which 
has significant impact on the interpretation of the CMD.  The crowding in 
the ground based images resulted in the Red Giant Branch (RGB) population 
being misidentified as the Red Super Giant (RSG) population.

Here we show how the details of the SFH of the last few Gyr obtained from 
uncrowded WFPC2 imaging helps us understand the properties of this galaxy, 
and why we believe the variable stars that H94 identified are in fact 
W~Virginis (W~Vir) or Anomalous Cepheid (AC) variable stars.

\section{Observations}

The HST data in this paper comes from the Wide Field Planetary Camera 2 
(WFPC2), using only 4 orbits per galaxy to obtain observations in three 
filters for a sample of nearby dwarf irregular galaxies (Skillman 1998).  
The sample consists of Sextans~A, Pegasus, Leo~A and GR~8.  The results 
have been quite dramatic and illustrate the tremendous advances possible 
even with short exposures when the problems of crowding have been virtually 
illuminated (Dohm-Palmer {\it et al.} 1997a,b, 1998; Tolstoy 1997; 
Gallagher {\it et al.} 1998).

Our data consist of HST/WFPC2 images, which cover the majority of the 
central region this small galaxy (see Figure~\ref{POSS}).  There are three 
exposures in the F555W~(V) and F814W~(I) filters (3$\times$ 600sec), and 
four exposures in the F439W~(B) filter (2$\times$900~sec and 2$\times$ 
1100~sec).  Each image is offset by a few pixels plus a fractional pixel 
offset, or ``dithered'' with respect to each other in an attempt to 
compensate for the under sampled PSF (Fruchter \& Hook 1996) and the intra 
pixel sensitivity variations (Biretta {\it et al.}  1996).  The images in 
each filter were combined by re-registering each image to nearest integer 
pixel, they were then cosmic-ray cleaned and combined using techniques 
described by Saha {\it et al.} (1996a).  The resulting highly resolved 
WFPC2 images of the heart of Leo~A are shown in Figure~\ref{MOS}.  We did 
not make use of the drizzle techniques because there were too few separate 
frames, and it was unclear what the noise properties, and hence the 
photometric fidelity, of the final images would be (see discussion in 
Dohm-Palmer {\it et al.} (1997a).

Ground based images were also taken with the Michigan-Dartmouth-MIT (MDM) 
Hiltner 2.5m telescope on Kitt Peak for calibration check on the HST V and 
I magnitudes.  The V (600~secs) and I(600~secs) images were taken on 23rd 
January 1998, in photometric conditions and 0.9~arcsec seeing.  The I 
calibration used 30 stars, and the fit rms was 0.013 mag (the average 
errors on the standards was 0.004), for V-I there were 28 stars, and the 
rms was 0.016 mag.  The calibration is reliable at least 2-3\% level, 
unless something very peculiar occured for which there is no evidence.  
These MDM images also cover a much larger area (9$\times$9 arcmin) than the 
HST images, and so we are also able to usefully increase the statistics for 
the brighter stars, such as those on the RGB.

\subsection{Colour-Magnitude Diagrams}

Photometry was carried out using a version of DoPHOT (Schechter, Mateo, \& 
Saha 1993) altered to to take account of a variable point spread function 
over each image (Saha {\it et al.} 1996a).  The HST photometry was 
calibrated and converted to the ``standard'' UBVRI system using the 
precepts laid out in Holtzmann {\it et al.} (1995).  This includes a 
correction to a 0.5~arcsec aperture.  The aperture correction for 
each filter and each chip and the accompanying errors are listed in 
Table~1. Also listed in Table~1 are the zero point (Zpt) and the
colour terms (T$_1$ and T$_2$) used in the photometric solution. 
These, and their 
errors come from the
Holtzmann {\it et al.} 1995 paper. Here are the photometric solutions
that we applied to our WFPC2 data to calibrate them on the BVI
standard system:

$$ B = -2.5\log (DN^{-1}, F439W) + [T_1 * (B-V)] + [T_2 * (B-V)^2] + Zpt
+2.5*\log GR $$

$$ V = -2.5\log (DN^{-1}, F555W) + [T_1 * (V-I)] + [T_2 * (V-I)^2] + Zpt
+2.5*\log GR $$

$$ I = -2.5\log (DN^{-1}, F814W) + [T_1 * (V-I)] + [T_2 * (V-I)^2] + Zpt
+2.5\log GR $$

\noindent{the } T$_1$ and T$_2$ terms in the above equations
have to be determined iteratively as the B-V or V-I colour of a
star is not a priori known. 
The $+2.5*\log GR$ is a term which is applied if the gain is not 14,
which is the value for the data which Holtzmann used to determine the
photometric solutions. It is +0.602 for our observations, at gain=7.

The photometry for F814W (I) and F439W (B) images were matched to the F555W 
(V) photometry, which was the most complete down to the faintest magnitude.  
The resulting CMDs are presented in Figure~\ref{CHIP1} and 
Figure~\ref{CHIP2} for the individual WFPC2 CCDs and for the combined data 
in Figure~\ref{CMD1} and Figure~\ref{CMD2}.  Figure~\ref{ERR} contains the 
error distributions which are critical in determining the quality of model 
fits to these data.

The HST CMDs for Leo~A contain several distinctive features.  There is a 
red clump (RC) between $23.  < M_I < 24 $, with V$-$I$\sim +0.8$.  Careful 
modeling is required to tell if this is truly an ``evolved'' RC, or if it 
is wholly, or some mixture with, the base of low metallicity helium-burning 
blue loop (BL) Stellar tracks.  The observed CMD also has a very pronounced 
BL Sequences, as seen in Sextans A (Dohm-Palmer {\it et al.} 1997a,b).  We 
cannot rule out the presence of either a red or a blue horizontal branch 
(HB, BHB).  The region expected to be populated by a BHB (I$\sim$ 24.7) is 
completely confused by the young stellar population along the Main Sequence 
(MS) and not accurately reached in V, B$-$V.

Looking at Figure~\ref{RDP1} it can be seen that the population mixture of 
the stars in Leo~A does not appear to vary significantly with position in 
the galaxy (also in MDM, see Figure~\ref{RDP2}).  The young (blue) stars 
are more centrally clumped, but the RGB population is not a smooth 
underlying population as is usually the case (i.e.  ``Baade's Sheet'') for 
a very old halo population, it is quite clumpy and not that much more 
extended out from the centre of the galaxy than the younger population.  In 
the case of Pegasus (Gallagher {\it et al.} 1998), there was a clear 
separation between the predominantly older population in the outer regions 
(their chip WF3, Figure~4) and the other chips closer to the centre of 
Pegasus.  In Leo~A the chip predominantly away from the central regions 
(WF2) still shows the presence of a MS and BLs, although we have only small 
number statistics.  This suggests that Leo~A is a predominantly young 
galaxy because there has not yet been time for a segregation of different 
age populations.  It is also possible that this segregation becomes less 
clear at the very low metallicity of Leo~A where the reddening of the 
stellar population due to age is very slow, and thus it is difficult to 
distinguish an old population (5$-$10~Gyr) from an intermediate age 
population (2$-$3~Gyr) along the RGB.

\subsection{Photometric Calibration Uncertainties of WFPC2 for Leo~A}

In Figure~\ref{MDMCMD} we show the MDM (I,~V$-$I) CMD for the whole field 
area, which clearly affords a better sampling of the younger brighter 
populations, such as the RGB, obviously seen in the well defined tip.  We 
can also see the serious effects of increased crowding in the ground based 
images.  The MDM RGB is much broader, and bluer than the HST RGB.  This is 
not a broadening of the RGB due to the inclusion of other regions of Leo~A 
with different RGB properties, but an intrinsic feature over the whole 
ground based image.  We have determined this to be
primarily an effect of crowding.  Stars at the red edge of 
the RGB can only be confused with stars which are bluer, hence making the 
trend of crowding to broaden the RGB in the blueward direction.  We
checked this by ``degrading'' HST images in V and I and then re-doing
the photometry. The result is an RGB that is broadened {\it exclusively}
to the red, and looks very similar to the MDM CMD in Fig~\ref{MDMCMD}. 

In Figure~\ref{MDMHST} we compare the absolute photometry of 
relatively uncrowded stars in the HST and the MDM images.  We select 
stars which have no neighbours within 10 pixels in the HST images, and in 
order that chip to chip effects are not a factor we do this on WF3 only.  
It could be that there is a systematic offset in colour between the HST 
calibration and the ground calibration, although the scatter is large 
(which could of course be due to problems with either calibration).  It is 
not possible to properly resolve this issue with these data, as there are 
too few stars which are uncrowded in the ground field to match the HST 
data. The simulations described in the previous paragraph suggest that the
offsets we see in Fig~\ref{MDMHST}
are primarily due to crowding in the ground based
images. Thus, despite efforts to pick only isolated stars,
we cannot tell with these data if there is any underlying
calibration offset of the WFPC2 versus standard UBVRI photometry.

\section{Basic Properties of Leo~A}

To properly model a CMD it is important to first decide upon the range of 
values that are acceptable for numerous basic parameters that affect the 
CMD models.  The most significant are the distance, the extinction and the 
metallicity distribution of a galaxy.  These properties can be determined 
independently from the CMD, and these independent measures must be 
consistent with the findings in a CMD.  These three basic parameters, in 
conjunction with observational errors and incompleteness (Tolstoy 1996) 
make the most significant impact on the properties of the CMD and hence the 
final SFH model.  The basic physical properties of Leo~A are summarized in 
Table~2.

\subsection{The Distance of Leo~A}

Unfortunately Leo~A does not have a reliable primary distance determination.
We compare several different indicators, summarized in Table~3, and
describe the properties below.

\subsubsection{Variable Stars in Leo~A}

The distance to Leo~A was thought to be well determined from the detection 
of 5 $\delta$-Cephei variable stars (H94).  However, recent observations of 
$\delta$-Cepheids in other galaxies (e.g., Pegasus, Aparicio 1994; IC~10, 
Saha {\it et al.} 1996b) call into doubt the reliability of using 
$\delta$-Cephei variable stars detected in only one filter.  From the WFPC2 
CMDs (see Figure~\ref{CMD1} and \ref{CMD2}), it is clear that the H94 Leo~A 
distance cannot be correct.  There is no way to reconcile either shape of 
the BLs or the presence of a RC with this large distance.  However, all but 
one of the $\delta$-Cepheids identified by H94 in Leo~A have roughly the 
colour expected for a bona fide $\delta$-Cepheid (Tolstoy 1996), and their 
light curves closely resemble those expected of $\delta$-Cephei variable 
stars.  The Tolstoy~(1996) colours came from observations taken on 
different nights, and so could only be shown to be broadly ``red'' or 
``blue''.  From these HST data it is possible to measure the colours for 3 
Cepheids that happen to fall into the WFPC2 field of view and the 
observations were taken within a few hours of each other (see 
Table~4).  On careful examination it can be seen that in fact two 
of the Cepheids are on the extreme blue edge of the classical instability 
strip, and the one which actually falls well in the classical instability 
strip was found by H94 to be too bright for its period compared to the 
others, and suggested to be an overtone pulsator.  Although we do not have 
a measure of $<$V$>$, the mean magnitude over the light curve, there is 
only a single V observation for these stars, so there is some uncertainty 
in where they exactly lie.  A possible explanation is that Cepheid (No.  9, 
H94) is the only bona fide $\delta$-Cepheid in the sample, and the other 
``Cepheids'' have colours more compatible with W~Vir stars or ACs or are 
not variables at all.  Using only one ``classical'' $\delta$-Cepheid makes 
the distance determination much less certain because of the intrinsic width 
of the instability strip (e.g., Tolstoy {\it et al.} 1995).  However, it 
does bring the galaxy much closer (m$-$M=24.9$\pm0.35$), although this 
value can be off by as much as 0.7~magnitudes in either direction due to 
the intrinsic width of the instability strip combined with only having a 
single $\delta$-Cepheid.

If we apply the W~Vir PL relation (which is not a very well established, 
see Appendix A), then we can use the single epoch WFPC2 V magnitudes to 
estimate a distance of m$-$M$\sim$24.5$\pm 0.5$ for Leo~A.  Despite the 
uncertainties in both the calibration, and having only one observation for 
each star (rather than $<V>$) which can affect the distance determination 
by as much as $\pm$0.5mag, we obtain a distance which is broadly consistent 
with that from the $\delta$-Cepheid, and the features seen in the HST CMDs.

If we have found W~Vir stars, this provides something of a dilemma, because 
they are typically very old stars usually seen in conjunction with BHB.  It 
is possible that, at this very low metallicity the BLs are blue enough at a 
low luminosity to enter into this region of the instability strip as young 
stars.  It is also possible that the BHB is hidden in the MS dominated by a 
young population.  If we assume that these variables are in fact ACs, which 
makes the non-detection of a HB less of a problem, then it would follow 
that H94 misidentified the true periods of this variables.  ACs typically 
have very short periods ($<$3 days).  As in Pegasus, these objects could 
also have much longer periods, of order hundreds of days.  However, the 
blue colours do not match with the properties of the long period variables 
found in other galaxies, which were extremely red (e.g., Tolstoy {\it et al.}  
1995; Saha {\it et al.} 1996b).  These variables in Leo~A could also be 
eclipsing binaries misidentified by H94.

\subsubsection{The Tip of the Red Giant Branch in Leo~A}

Determining the distance to a galaxy using the tip of the red giant branch 
(TRGB) is only reliable for galaxies with a predominantly old population 
($>$2~Gyr), and a moderate metallicity.  For metallicities lower than 2\% 
solar, assuming efficient convective overshooting, the upper mass limit for 
helium flash is only about 1.6 $-$ 1.7 $\Msun$, and so the RGB phase 
transition only occurs after 1.3 $-$ 1.5 Gyr.  The luminosity of the TRGB 
is thus a strong function of age in this regime (Sweigart, Greggio \& 
Renzini 1990) and so it cannot be reliably used as a distance indicator.  
The efficacy of the TRGB as a distance measure is also somewhat uncertain 
in star forming, low metallicity systems such as this, because the AGB 
population (if one exists) is distributed {\it above} the true TRGB without 
the characteristic down turn of higher metallicity AGB stars.  Thus, 
depending upon the star formation history of the galaxy this could 
effectively mask the true tip.  

The HST TRGB in Leo~A is very indistinct due to the small number of stars 
populating it.  The MDM field covers a much larger area and hence we get a 
more accurate and clear TRGB measurement of m$-$M=24.5 $\pm$ 0.2 (see 
Figure~\ref{MDMCMD}), assuming M$_I~=~-$4., from the calibration of Lee, 
Freedman \& Madore (1993), and that crowding isn't a significant problem 
(e.g., Martinez-Delgado \& Aparicio~1997). The presence of a clear TRGB
in Figure~\ref{MDMCMD} is an indicator that there is an old population
in Leo~A (older than 1$-$2~Gyr). As we show later (in \S 4.3) however,
the ratio of numbers of RC/RGB stars does not allow that this older
population can be very large.

In Figure~\ref{RGB} we have over plotted observed RGB sequences of several 
Galactic globular clusters (from Da~Costa \& Armandroff 1990) for 
comparison with the observed Leo~A RGB.  This shows how varying the assumed 
distance effects the age and metallicity determination of the system with 
very little effect on the TRGB magnitude.  All these values have to be 
consistent with what we know of Leo~A.  At m$-$M=24.6 we either have to 
assume that the stars are not globular cluster age, or that they are at an 
even lower metallicity than M~15, which, at [Fe/H]=$-$2.2, is one of the 
lowest metallicity globular clusters known.  At m$-$M=24.2 we can find 
stars which are consistent with an extremely metal poor ([Fe/H]=$-$1.9) 
globular cluster age population in Leo~A.


\subsubsection{The Leo~A Red Clump}

It has recently been shown that a well defined 
RC can provide a useful distance
determination (Cole 1998; Paczy\'{n}ski \& Stanek 1998).
RC stars are in the core helium-burning evolutionary phases, and their 
luminosity varies depending upon age, metallicity and mass loss (e.g., 
Caputo, Castellani \& Degl'Innocenti 1995).  The RC is relatively blue and 
thus, consistent with the upper limits which can be assumed from the young 
population, we conclude that we see an extremely metal poor RC 
(Z$\sim$0.0004). We also note that this RC 
has a large extent in magnitude, $\sim$1.1~mag in V.  
We can use this extent to estimate the age of the population that produced 
it using theoretical models (Caputo {\it et al.}  1995; Girardi, Bressan \& 
Bertelli 1998, in preparation).  This age measure
is {\it independent of absolute magnitude and hence distance}.

The luminosity of the RC was computed from the I luminosity function 
between 0.65 $<$ V$-$I $<$ 0.95 and 22 $<$ I $<$ 25.  The underlying RGB 
stars were removed using a linear fit to the wings of the I histogram, and 
fitting the residuals ($\sim$2400 stars) to a Gaussian.  Added weight was 
given to the peak I mag of the fitted Gaussian profile and the 
low-luminosity side of the histogram, because the high side is obviously 
contaminated by faint BL stars.  Thus, the average RC is at I = 23.77, 
(V$-$I) = 0.82.  Castellani \& Degl'Innocenti (1995) computed isochrones 
for Z = 0.0001 and Z = 0.0004 for 0.7 to 20 Gyr populations, and also made 
some RC models for the same metallicities (Caputo, Castellani \& 
Degl'Innocenti 1995), and by using the total V extent of the RC, their 
models predict an upper limit to the age of the population assuming 
M$\rm_I$ = $-$0.4 for the RC.  The Gaussian profile fitted to the I data 
has a width of 1.0 magnitude at 10\% of the peak (or 1.1 mag in V).  
Assuming the youngest RC possible (1$-$2~Gyr) and a reddening, A$\rm_I$ = 
0.04, the distance modulus, m$-$M=24.2 $\pm$0.2.  If the RC is 
predominantly at its oldest extreme (9$-$10~Gyr) then the distance of Leo~A 
is 23.9 $\pm$0.1, but this distance doesn't fit in with the BL or the TRGB.

Thus, the mean I magnitude by itself does not rule out an older RC, but the 
vertical extent does. This
cannot be reproduced for ages older than a maximum of 
$\sim$2.2 Gyr.  A 1.5$-$2.2~Gyr old RC, will have M$_I \sim -0.4 \pm 0.1$, 
which gives a distance of M$-$m = 24.2$\pm$0.2.  For older (4.5$-$10 Gyr) 
and younger (0.8$-$1.5 Gyr) models, the RC is fainter, possibly as faint as 
I = 0.0 $\pm$ 0.1, but these latter two possibilities are in conflict with 
the TRGB distance, for the old models, and with the lack of MS, for the 
young models. And with the BLs, in both cases.  For an example of a RC 
clearly older than 2 Gyr, see the Hipparcos solar neighborhood CMD 
(Jimenez, Flynn \& Kotoneva 1998).

\subsubsection{The He burning ``Blue Loops'' in Leo~A}

Despite the uncertainty involved in BL physics we note that the Sextans~A 
HST data shows excellent fits to the BL models (Dohm-Palmer {\it et al.} 
1997b).  We also note that the best fitting distance to the BL models for 
Leo~A appear able to match our data very well, at Z=0.0004, they give m$-$M= 
24.1$\pm0.1$.  Because of the large uncertainties involved in creating 
theoretical models of BLs (e.g.  Meynet 1992, Chiosi 1997) they are not 
generally very reliable distance indicators.  There are several case where 
the MS and the BL shapes are clearly inconsistent, e.g.  in Pegasus 
(Gallagher {\it et al.}  1998).

\subsubsection{A New Distance to Leo~A}

Because we need to know an accurate distance to Leo~A to interpret the CMD 
in terms of a SFH we chose the distance (from Table~3) which is 
consistent with the largest number of stars in the CMD, namely the stars in 
the RC, rather than the TRGB (a more typical distance measure, although 
there are strong caveats for galaxies like Leo~A).  The RC as well as the 
luminosity (and shape) of the BLs favour a distance modulus around 24.2.  
This distance is also broadly consistent with the (low end) of the variable 
star distance determinations and the TRGB.  Thus, based on this new HST 
CMD, we assume the new distance to Leo~A to be m$-$M=24.2$\pm$0.2.  Where 
the error is purely the range that allows a reasonable fit to the CMD 
models.

\subsection{Extinction Towards and Within Leo~A}

Burstein \& Heiles (1984) give a low extinction of E(B$-$V)=0.02 towards 
Leo~A.  Following the approach we used for the Pegasus dwarf (Gallagher 
{\it et al.} 1998), we checked the region for excess 100~$\mu$m far 
infrared emission in the IRAS map; none was found which is consistent with 
the Burstein \& Heiles estimate of the extinction.  HI synthesis maps of 
Leo~A by Young \& Lo (1996) show peak internal column densities of 
$<$3$\times$10$^{21}$~cm$^2$.  For an SMC extinction-to-gas ratio, the 
maximum internal extinction is predicted to be E(B$-$V)$<$0.03 (Fitzpatrick 
1985).  Additional support for a low value of the extinction comes from the 
spectrophotometry of a planetary nebula in Leo~A by Skillman, Kennicutt, \& 
Hodge (1989).  They found an anomalously low ratio of H$\alpha$ to H$\beta$ 
emission line intensities in their data, and thus see no evidence for 
extinction.  We adopt as baseline values A$_B=$0.08, A$_V=$0.06, and 
A$_I=$0.02 in our subsequent analysis of the Leo~A CMDs.

There is circumstantial evidence from looking at the properties of 
background galaxies seen through Leo~A that there maybe be variable 
internal extinction within the central regions of Leo~A.  Figure~\ref{EXT} 
shows a sub-section of the WF3 chip containing two galaxies very close 
together which seemingly have quite different extinction properties.  This 
of course maybe due to some intrinsic difference in the properties of these 
galaxies - although typically background galaxies are to first order fairly 
similar (Tyson 1988; Lilly, Cowie \& Gardner 1991).  So this is not 
conclusive proof (a more careful study is in progress, Tolstoy, Freudling 
\& Rosa, in prep), merely suggestive, and it fits in with other evidence 
for variable extinction seen in our CMDs.  The scatter of stars between the 
MS and BL at I$\sim$22.5 and further down the MS (to the red) may be the 
result of variable extinction through the star forming regions in the 
centre of Leo~A.  The distribution of these points is inconsistent with 
MSTOs for example.  However, the global reddening towards Leo~A seems to be 
correct because the MS lies where it should (unlike in Pegasus, Gallagher 
{\it et al.} 1998).  Variable extinction may be common in irregular 
galaxies, and, for example, has been clearly identified in multi-colour 
data in the area around SN1987A in the LMC (Romaniello {\it et al.} 1998).

\subsection{The Current Metallicity of Leo~A}

The only metallicity determination for Leo~A comes from a PN (Skillman, 
Kennicutt \& Hodge 1989), which happens to be the brightest H$\alpha$ 
source in the galaxy.  The metallicity was found to be extraordinarily low, 
12+log(O/H)=7.3$\pm$0.2 ($\sim2.4$\% solar), for a gas rich star-forming 
galaxy.

The young population, specifically BLs, are the most sensitive global 
feature in a CMD to metallicity.  They do not require an assumed age as the 
RGB does.  The BL in our CMD are consistent with Padua (Fagotto {\it et 
al.} 1994) Z=0.0004 (2\% solar), and not with Geneva (Schaller {\it et al.} 
1992) Z=0.001 (5\% solar) for any reasonable distance.  This low value is 
also consistent with the Skillman {\it et al.} PN metallicity, but this 
galaxy is desperately in need of more detailed abundance work.

\section{Making Models}

It is important to understand the theoretical stellar evolution and 
atmosphere models being used and their limitations.  Many different aspects 
of this are discussed in detail in Tolstoy~(1996), Tolstoy \& Saha~(1996) 
and Gallagher {\it et al.}~(1998).  Errors in the models have very serious 
implications for Leo~A because of the insensitivity of low metallicity 
models to age.  The RGB isochrones crowd together in a very limited 
parameter space, and so small changes in stellar physics can change an 
optimum model from a young 2~Gyr old population to an old 10~Gyr 
population.

For all the modeling presented in this paper we have used either Z=0.001 
Geneva stellar evolution models (Schaller {\it et al.} 1992), or, most 
extensively, Z=0.0004 Padua stellar evolution models (Fagotto {\it et al.} 
1994).  We do not like to ``mix'' models like this, and we note that when 
comparisons have been possible Geneva and Padua give different results for 
identical initial conditions, and Padua tends to produce a redder RGB than 
Geneva (see Figure~11 in Gallagher {\it et al.}~1998).  In all cases we 
have used the standard Kurucz~(1991) models to convert the temperature and 
luminosities of the stellar evolution models into the observed colours in a 
CMD.

The Tolstoy~(1996) distance and metallicity clearly don't work for this new 
more accurate data.  It is relatively easy to mistake RGB for RSG stars if 
the CMD is not sufficiently detailed to detect any other features (e.g., RC 
or HB).  The poor match between new data and models isn't merely a problem 
of distance but also metallicity.  Changing the distance alone means that 
the RC is too faint and too red, and the RGB is also too red to ever match 
the observed data.

The other important considerations in the construction of CMD models are 
Initial Mass Function (IMF) and incompleteness.  We treat them in the same 
way as discussed in Tolstoy~(1996) and Tolstoy \& Saha~(1996).  Namely, we 
use a Salpeter IMF throughout.  We are completely insensitive to any 
possible variations in the IMF, because we are predominantly sampling stars 
with masses $>1\Msun$.  
In this HST data incompleteness due to crowding is 
not a problem (see Fig~\ref{MOS}).  This was checked on more crowded fields 
of Sextans~A by Dohm-Palmer {\it et al.}  (1998a).  
The incompleteness basically 
follows the photometric error distributions computed by DoPHOT. 
There is a much sharper cut-off at faint magnitudes than is typical for 
crowded fields.

We use the methods laid out in Tolstoy \& Saha~(1996) to compare our data 
with possible models and find the most likely model to match the 
observations.  When we display our models we convolve the model with an 
error distribution to make it easier for the eye to see what is a good 
match to the data.  This error distribution is taken straight from the 
measurement errors (shown in Fig~\ref{ERR}) and randomly added to each 
filter separately in the model.  However, as explained in Tolstoy \& Saha 
(1997), one of the strengths of this method of comparing data and models is 
that we compare them numerically using the measurement error that comes 
with every data point as a probability density to predict the probability of 
a model being consistent with this measured point.  In each case our 
favoured model is one which maximizes the Likelihood that the distribution 
of model points matches that of the data points within the observational 
errors on the data points.  This of course assumes that a model is 
``perfect'', which for our purposes it is.  It is not clear how to give 
theoretical models errors.  However were they provided we could easily 
incorporate them into our modelling code.

We split up our analysis of the CMD into two different areas, namely the 
``young'' population (represented by the MS and BL stars, thus in the case 
of these observations back to $\sim$1~Gyr ago) and the intermediate/old 
populations seen in the RC and RGB ($\sim~1-$10~Gyr).  The approaches for 
these regimes are different, although we do not hold them entirely separate 
in the modeling process, because the contribution of the young BL stars 
overlying the older RC is important.

\subsection{Metallicity Evolution}

Accounting for metallicity evolution in a CMD model is difficult.  It is 
impossible to determine a unique model based solely on the RGB.  This is 
because the variation of stellar evolution tracks with metal levels and 
distributions is not necessarily linear; as a result simple interpolations 
between tracks with widely-spaced metallicities can be misleading.  
However, if metallicity evolution is neglected in a CMD model then the best 
model for that galaxy will be younger than if metallicity evolution were 
included.

When a galaxy makes stars, then the detritus of this process (e.g., from SN 
explosions and stellar winds) make it unlikely that the galaxy can avoid 
metallicity evolution all together (e.g., Eggen, Lynden-Bell \& Sandage 
1962).  However, there is no concrete observational evidence that this is 
true, although abundance ratios of different elements do give us model 
dependent suggestions (e.g., Pagel 1994).  In the disc of our Galaxy, for 
example, it was recently shown that, although there is a general trend in 
metallicity evolution with time, the scatter is always large (Edvardsson 
{\it et al.} 1993).  We do not understand in detail how stars interact with 
their surrounding ISM, and thus how current star formation feeds the metal 
enrichment of future generations.  Looking at recent results of absorption 
line studies of Zinc abundances at cosmological distances ($z = 0.7 - 3.4$) 
there is again evidence for a shallow evolution of metallicity in galaxies 
over this long redshift range, but there is also a large scatter in values 
at any time (Pettini {\it et al.} 1997).  Absorption line studies of these 
species provides arguably the most reliable estimator of the metallicity of 
the {\it gas} in a galaxy.  If suitable background continuum sources could 
be found behind nearby galaxies this would dramatically improve our 
understanding of how the ISM in different galaxies evolves and is affected 
by the proximity of current star formation.

However, in the absence of these probes we are forced to rely upon the 
emission line measure of HII regions and PN.  While the reliability of 
these measurements has been questioned due to concerns of ``pollution'' of 
these regions by the products of massive star evolution (Kunth \& Sargent 
1986), recent studies have found no evidence for this effect (e.g., 
Kobulnicky \& Skillman 1997 and references therein).  In addition to 
providing measurements of the present ISM metallicity, these emission lines 
studies can provide hints about the past chemical enrichment of the 
surrounding ISM (e.g., Skillman, Bomans \& Kobulnicky 1997).  Models can be 
made from emission line ratios (e.g., C/O and N/O) which indicate a 
possible global secular chemical enrichment history for a galaxy (e.g., 
Kobulnicky \& Skillman 1998) if it can be assumed that HII regions are 
representative of the global environment, as discussed at length in 
Kobulnicky \& Skillman (1998).

At the low metallicity of Leo~A, the age-metallicity degeneracy is 
particularly acute (Da~Costa 1997), and so only sensitive to gross changes 
in assumptions.  Another practical problem with fitting any kind of 
metallicity evolution in Leo~A is that for the young population we are 
already using the lowest metallicity set of consistent model stellar 
evolution tracks available and so there is no parameter space left to model 
more metal poor stars.  There is even some argument about what is the 
lowest metallicity that an object can have and with Leo~A we are pretty 
close to perceived limits for galaxies (Kunth \& Sargent 1986).

Even though it would be desirable to include metallicity evolution it is 
not clear how to do this.  If we interpolate between different metallicity 
tracks we are going through very uncertain parameter space.  Metallicity is 
not a parameter in a star which can be so easily described.  The only way 
to be able to do this properly is to have a much finer grid of models at 
different metallicities.  This is not yet available.  Thus we opt not to 
include the effects of metallicity evolution in our modeling, but to 
mention how a further decrease in the metallicity going back in time would 
alter our conclusions.

\subsection{The ``Young'' Population: The
Main Sequence and the He Burning ``Blue Loop'' Stars}

We know that Leo~A is currently forming stars.  It has several faint HII 
regions, and we see a distinct MS and BL stars in the CMDs (see 
Figure~\ref{CMD2}).  Comparing our observed CMDs with theoretical stellar 
evolution models we can determine the recent SFH of this galaxy.

First of all we determine the evolutionary time scales to which we are most 
sensitive.  The MS luminosity distribution means that we can detect MSTOs 
between 800~Myr old (we are restricted from going further back in age by 
the sensitivity limits of our data) and $\sim$10~Myr, which is our 
saturation limit and means that stars younger than this are too bright for 
our detector.  Although we are sensitive to star formation at all times 
along the MS, we are not able to characterize it well because we cannot 
detect the most massive turnoff stars.  Towards the base of the MS we 
become less and less sensitive to anything but large variations in the SFR.  
This is partly due to increasing error and incompleteness uncertainties, 
but also due to the fact that any variations in SFR further back in time 
are ``hidden'' by subsequent star formation going on until the present 
overlying the entire MS.

The BL colour and luminosity are able to give us detailed SFR information 
back to just beyond a Gyr ago.  The BLs are more luminous than the stars of 
the same age on the MS and hence the BL is valuable in giving us accurate 
information to feed into MS models.  Subsequent epoches of star formation 
do not overlie each other on the BLs (unlike on the MS).  However, after 
about 900~Myr the BL and the RC stars merge together at this metallicity, 
and it is difficult to accurately disentangle them.  Unfortunately, in the 
current data, the MS is not deep enough to help disentangle the BL/RC 
confusion uniquely.

Another feature of a very young population are the RSGs which have a 
vertical sequence extending above and slightly blueward of the RGB.  These 
are very young stars corresponding to roughly the ages and masses of the BL 
stars at the same luminosity.

\noindent{Our} detailed modeling of the young population along the MS 
and BLs of Leo~A within the above limits results in Figure~\ref{RDP3} 
and the following conclusions:

\begin{itemize}

\item{}The gap in the BL sequence
(I$\sim$23., V$-$I$\sim$0.4) combined with the relative numbers on the 
upper and lower main sequence point to a gap in star formation between 
500$-$900Myr ago.  The 500Myr and 900Myr limits come from the blue and red 
extents of the gap.

\item{}The relative numbers of stars on the MS and BLs are consistent
with a more or less constant SFR over the last 500Myr.  There are 
variations in the stellar number densities which suggest that this has not 
been completely smooth, although there is no evidence for more than a 
factor around 2 variation over this time period.

\item{}The spread of stars at the bottom of the CMD is
inconsistent purely with error scatter, although it is very easy to 
underestimate the errors and the incompleteness of data close to the 
observational limits especially in WFPC2 data with the warm pixel problems 
(cf.  Dohm-Palmer {\it et al.} 1997a).  An active period of star formation 
about 900Myr ago (consistent with the BL gap) would have MSTOs populating 
this area of the diagram (see Figure~\ref{YG}).  We are limited by the 
implications such a turnoff has for the number of stars predicted higher up 
in the RC/RGB.  But it is likely that some stars from lower (older) MSTOs 
to which we are not directly sensitive are more likely to be up scattered 
and thus produce a higher number of stars at the younger turnoff at the 
edge of our observational limits.

\end{itemize}

\subsection{The ``Intermediate'' Age Population: The Red Clump and the
Red Giant Branch}

The dominant population of red stars in Leo~A resides in a RC 
(N(RC)~$\sim$~2400 stars in I,~V$-$I), which has already been discussed in 
\S~3.1.3.  There is also a relatively weak RGB (N(RGB)~$\sim$~570 stars in 
I,~V$-$I) extending above the RC.  The ratio of the numbers of stars in 
these two phases [N(RC)/N(RGB)] is very sensitive to the SFH of the galaxy.  
The classical RC and RGB appear in a population at about the same time 
($\sim$ 0.9--1.5 Gyr, depending on model details), where the RGB are the 
progenitors of the RC stars.  The feature that ties these phases together 
is the explosive ignition of helium under degenerate conditions at the RGB 
tip (the He-flash).  There is a RC, therefore there was a He-flash.  This 
results in a near-constant M$_{core}$ for RC stars 
between $1.7 < M_{star} < 
0.8 \Msun$ and hence near-constant RC lifetime, t$_{RC}$.  The age of onset 
of the He-flash is between 0.9 and 1.5~Gyr.  Regardless of the absolute 
time scale the ratio of N(RC)/N(RGB) in Leo~A is a firm indication that the 
galaxy is right around the age of He-flash onset.  But this isn't a very 
precise indicator because the RGB lifetimes depend on stellar mass and 
because of mass-loss there is no unique mapping from RGB to RC mass.  
Massive RGB stars build core mass more quickly than their low-mass 
counterparts and thus achieve helium ignition in a short amount of time; 
t$_{RGB}$ is a strongly decreasing function of M$_{star}$ Hence the ratio, 
t$_{RC}$ / t$_{RGB}$, is a decreasing function of the age of the dominant 
stellar population in a galaxy.  Thus, the higher the ratio, N(RC)/N(RGB), 
the younger the dominant stellar population in a galaxy.  For Leo~A the 
N(RC)/N(RGB)~$=$~4.4, which suggests that the RC is very young.  A 
predominantly much older model contains either no RC ($>$9~Gyr) or an under 
luminous one (2$-$9~Gyr), as shown in Figure~\ref{INT}, where this ratio 
tends to $\sim$1.  In Pegasus we see a slightly older (and more metal rich) 
RC, where N(RC)/N(RGB)~$\sim$~2 (Gallagher {\it et al.}  1998), 
which suggests a 
RC age of 4$-$5~Gyr.

For a given metallicity the RGB red and blue limits are given by the young 
and old limits of the stars populating it.  For Z=0.0004 the youngest stars 
to populate the RGB (rather than the Super giant regions) are $\sim$1~Gyr.  
As a stellar population ages the RGB moves to the red, for constant 
metallicity, the red edge is determined by the age of the oldest stars.  If 
there is metallicity evolution, then the age-metallicity degeneracy makes 
it impossible to disentangle these two effects on the basis of the optical 
colours of the RGB.

The RGB below the RC also gives some limits to the older MSTO ages.  
Although these limits are fairly similar to those on the MS, they provide a 
useful consistency test for assumptions made about a burst of star 
formation ending 0.9~Gyr ago.

\noindent{Our} detailed modeling of the RC in Leo~A leads to the following 
conclusions:

\begin{itemize}

\item{}The optimum 
model for the number of stars in the RC relative to the RGB and the 
luminosity distribution of the clump itself, is one in which the RC formed 
in a short ``burst'' of star formation over the period 900$-$1500~Myr ago.  
This fits in with the young stellar population models based on the MS and 
BLs.  If the RC models are accurate it is not possible to match 
significantly older RC to the data.  The RGB however matches only much 
older models (9$-$10~Gyr).

\item{}A young model does have problems, the most serious of which
that any successful RC model makes the young RGB around 0.2 mag too blue in 
V$-$I at the TRGB and a decreasing offset down to zero at the RC.  This 
could be due to a problem with the stellar physics used in the models.  It 
is well known that the red extent of the RGB is very sensitive to the 
assumed core mass on the RGB and hence to assumptions made about the 
evolution of the core.  Given that the Padua RGB is typically redder than 
the Geneva RGB for the same mass star at higher metallicities (e.g., 
Gallagher {\it et al.} 1998), it might be that Geneva tracks, using 
slightly different assumptions about stellar physics (e.g., different 
values of the $^{12}$C($\alpha$,$\gamma$)$^{16}$O cross section) would give 
a more consistent picture between RC and RGB ages.

\item{} There is a possible
AGB star population, as would be expected for a predominantly 0.9$-$1.5~Gyr 
old population, directly above the TRGB.  The Padua isochrones predict that 
AGB stars would lie directly about the TRGB at this metallicity, with no 
turnover to the red (Bertelli {\it et al.} 1994).  Some of the bright stars 
significantly red ward of the RGB may also be evolved AGB or carbon stars.  
Spectroscopy or infrared imaging would confirm this.

\end{itemize}

In summary, comparing our data with Padua 2\% of solar metallicity 
theoretical stellar evolution model tracks suggests that the RC is very 
young and the RGB is very old.  It is not clear that we can reconcile these 
two results with the numbers of RC and RGB stars in the data.  We prefer 
the young RC model which fits most of the data and results in a 
predominantly young galaxy.  Based on the young RC model at least 60\% of 
the stars ever formed in this galaxy were formed in a short burst of star 
formation about 900$-$1500~Myr ago.  In determining the length and 
intensity of this burst, we have to trade off the number of stars at 
different luminosities in the RC.  If we did not have a problem with the 
inferred colour of the RGB we could assume that the entire galaxy was less 
than a few Gyr old.

\subsection{Is There a Globular Cluster Age ``Old'' Population?}

The fact that the optimum RC model does not properly predict the observed 
RGB data may either suggest that the theoretical models are in error, or 
that there is an underlying much older population in Leo~A.  Exactly how 
much older is impossible to determine, without deeper photometry and 
information about the metallicity evolution within Leo~A.  The only 
indications of an old population in the present data are buried in the RGB, 
and in the HB branch region.  In star forming systems the HB is quite easy 
to ``lose'' in the MS and MSTOs.  The RGB similarly is overlaid by the 
number of different age populations and is impossible to disentangle 
because of the age-metallicity degeneracy.

\noindent{Our} exploration of the existence of a much older 
population in Leo~A results in the following conclusions:

\begin{itemize}

\item{}If the theoretical models and the HST calibration is to be believed,
then the observed RGB fits the 2\% Padua models for a 9$-$10~Gyr old 
population.  This old a population does not have a RC, although it does 
assume that a HB is hidden in the ``noise'' of the MSTOs in this star 
forming galaxy.  It is difficult to obtain even a limit for the presence 
(or lack) of a HB with the current data.  But, relying on the RC and HB for 
an indication of an old or intermediate age population is risky.  These 
relatively short-lived phases of evolution can show dramatic variations in 
their CMD morphology between SFH scenarios with bursts of star formation 
rather than uniform SFR over a long period of time

\item{}To avoid relying only on theoretical models we also compared our
data with observed globular cluster RGBs (Da~Costa \& Armandroff 1990) in 
Figure~\ref{RGB}.  The globular cluster RGB which best matches the Leo~A 
RGB is NGC~6397, which is consistent with a much older ($>$10~Gyr) and more 
metal poor ([Fe/H]=$-$1.9, 1.2\% solar) population.  If we do not wish to 
include significant metallicity evolution, and, indeed, a more metal poor 
galaxy than 2\% has never been observed in emission, then the observed RGB 
is too blue for the matching metallicity RGB ([Fe/H]=$-$1.7).  Hence Leo~A 
could be considerably younger than a globular cluster.

\item{}We cannot expect a large old population because the RGB is so
sparsely populated.  The modeling of the MS+BL+RC can account for at least 
90\% of the stars seen in the present CMD, so any globular cluster age 
population can represent only 10\% of the star formation in Leo~A.

\end{itemize}

To summarize, the theoretical models suggest we have an old population, at 
least 9$-$10~Gyr old.  This is a very uncertain number because of the small 
number of stars we detect in the RGB and the absence of a clear HB and our 
total lack of information about the metallicity evolution with time in this 
galaxy.  It is also possible that either the stellar evolution models, or 
the stellar atmosphere models converting temperature and luminosity to 
observed colours, are inaccurate in this low metallicity, young regime, and 
that the RGB is solely young.  This would be most consistent with the RC 
modeling.  Another more worrying possibility is that there is an error in 
the standard calibration of the HST filters.  Thus, we can neither rule out 
the presence of an old stellar component in Leo~A nor unequivocally prove 
there has to be one.

\subsection{An Optimally Matched Model for the Leo~A CMD}

The modeling presented in the last 3 sections results in the combined 
optimum model shown in Figure~\ref{MOD}.  There are two panels on this 
plot.  One shows the model in red, and the other splits up the different 
age components that make up the model ($<$500~Myr, 900$-$1500~Myr and 
9$-$10~Gyr) into three different colours.  This last plot shows up the 
different age stellar populations and where they contribute.  The following 
points are clear from Figure~\ref{MOD}:
\begin{itemize}

\item{}This model matches the observed MS effectively. 
The small scatter to the red side is presumed to be due to non-uniform 
reddening intrinsic to Leo~A, or to the presence of binary stars.

\item{} The RC is modeled convincingly, although no possible model for
the RC can match the RGB distribution.  This maybe be due to failings in 
the theoretical models used or to the HST calibration.

\item{}There are 
data not matched by the RC model which could not be helped by any extra BL 
or RC model (or mixture thereof).  This is probably due to underestimates 
of the errors or incompleteness at fainter red magnitudes, and perhaps also 
the effects of variable reddening and/or binary stars. There could
also be HB stars here, or faint BL stars, which some models 
(e.g. Girardi {\it et al.} 1998) predict to lie 0.1$-$0.3~mag below the RC.

\item{}The
RGB can be well matched only by the Padua models of very old ages.

\end{itemize}

Thus, we are confident that we have managed to match the major features in 
the CMD with these models.  It is by no means a unique model and the 
mismatch of the observed and model RGB is a cause of serious problems in 
finding the optimum model using the Tolstoy \& Saha (1996) methods.  The MS 
and BL make it obvious that there has been fairly active SFR over the last 
500Myr, and the RC makes it hard to avoid a fairly strong peak in the SFR 
0.9$-$1.5~Gyr ago.  The exact duration and intensity of these star 
formation periods are dependent on small number statistics.  If there were 
significant metallicity evolution the conclusions based on the RC would 
alter.  If the RC were more metal poor (e.g., Z=0.0001) than the MS then 
its maximum age would be older than that based on Z=0.0004 models, but only 
by a few hundred million years (Castellani \& Degl'Innocenti 1995).  In our 
optimum model in Figure~\ref{MOD} we also include an old population.  This 
population appears to fit the RGB stellar distribution very well, but it is 
strongly reliant on the stellar evolution and atmosphere models, which are 
not well tested in this regime.  This conclusion also assumes that our HST 
data has been accurately calibrated (see comparison with ground based data 
in \S2.1 and Figure~\ref{MDMHST}).

\section{A Proposed Star Formation History of Leo~A}

In Figure~\ref{SFH} we show the proposed SFH for Leo~A which created the 
optimum CMD model in Figure~\ref{MOD}.  We use pseudo units to describe out 
SFR because we have not calibrated these values onto an absolute scale.  
The SFH in Figure~\ref{SFH} is split into two parts with differing time 
resolution.  The error bars give a rough estimate of the reliability of a 
given SFR.  During the quiescent periods it gives an estimate of the 
intensity of SFR that would be difficult to hide.  This is a difficult 
error to make absolute.  We can hide a $<.0005$ pseudo SFR units star burst 
of short duration ($< 10^8$ yrs) between 2 and 10 Gyr quite easily.  But if 
there are too many of them, or one lasts too long then there will be 
problems for the model (the RGB will become over populated and perhaps 
structured).  The error bars make no allowance for errors in distance, 
reddening, metallicity or model inaccuracies.  These are the major error 
sources of uncertainty in creating this SFH.  If one of them is wrong the 
whole scenario presented here could change dramatically.  The error bars in 
Figure~\ref{SFH} are not directly related solely to the statistics of the 
number of stars involved.  It is very hard to simply put error bars down in 
the strictest meaning of an error bar.  If there is any variation in the 
number of stars in one `element' of the SFH this will affect other elements 
to maintain a good match to the data.

The younger regime in the upper panel of Figure~\ref{SFH} is that to which 
we are most sensitive to small variations in the SFR with time.  We have MS 
information over the last 800~Myr which gives the most accurate SFR measure 
versus time, although with decreasing accuracy with time from present due 
increasing errors and smaller number of stars from the older populations.  
The SFR over the MS is broadly consistent with a constant SFR over the last 
500~Myr.  Between 500 and 900 Myr the data are consistent with either zero 
or a much lower SFR in Leo~A.  This comes from the gap in the BLs clearly 
seen just blueward of the RC.  This may be a statistical effect in the 
sampling of the BLs, but it also ties in with our number density along the 
MS.

In the intermediate age range the RC gives us limits to the SFR that must 
have created it, but it is very insensitive to any variations in the SFR 
during this period.  Thus we know the RC contains stars older than 900~Myr 
but younger than 1500~Myr from it's luminosity range and the ratio 
N(RC)/N(RGB) stars, but we cannot be sure if the SFR was constant or 
varying between these limits.

On the bottom panel we show the old population seen in Figure~\ref{MOD}.  
This is highly speculative more than 2~Gyr in the past.  The long gap with 
apparently zero star formation represents the gap in the models between the 
young RGB (at 0.9$-$1.5~Gyr) and the old one (at 9$-$10~Gyr).  This gap 
could be made to disappear if it were found that the young RGB models were 
too blue or in the case of a calibration error.  Presuming this quiescent 
era in the galaxy is real it is possible that there were short bursts of 
star formation all through this period rather than the prolonged period at 
10~Gyr.  However this would require a steady metal enrichment during this 
period to keep the RGB so narrow.  We also must conserve the total number 
of stars found on the RGB so the amount of star formation possible $>$2~Gyr 
ago has a hard limit from the number of stars seen on the RGB.

One place that we can gain some insights into possible age-metallicity 
correlations in very small galaxies is from the 
Galactic dSph systems. While all of these objects currently contain neither 
cool gas nor young stars, they clearly were small, star-forming galaxies 
only a few Gyr ago (see Smecker-Hane {\it et al.} 1994, Stetson, Hesser, \& 
Smecker-Hane 1998).  Like most dwarf galaxies, mass and mean stellar 
metallicity roughly correlate with galaxy mass in dSph 
systems (see Gallagher \& Wyse 1994, Mateo 1998), but it is not clear 
that metallicity relates to stellar {\it ages}, particularly 
in the Carina dSph. The dSph therefore serve as warning that age-metallicity 
relationships are not necessarily simple, even in extreme dwarf galaxies 
and it is therefore premature to assume that we know the form of the 
age-metallicity relationship, if one exists at all, in Leo~A.

We note that, consistent with our CMD modeling, the unusual fraction of 
true/anomalous Cepheids hints that this galaxy must have had a higher SFR 
in the past than at present, and/or that the metallicity affects this 
ratio.  If we assume that the faint variables are young AC or W~Vir stars 
this suggests that the SFR in this galaxy was higher at some point around 
1$-$3~Gyr from the present.  The exact magnitude of past SFRs is difficult 
to predict from these observations alone because of the small number 
statistics and the incompleteness errors in monitoring these galaxies from 
the ground.  This is broadly consistent with our model based on the CMD 
properties.

On the basis of these new HST observations, H94 primarily detected some 
kind of W~Vir, or AC type of object, which are known to closely resemble 
their brighter cousins, but to be $\sim 2 - 2.5$ magnitudes fainter.  These 
are thought to be ``young'' single star W~Vir/AC stars due to the fact that 
the Leo~A CMD does not contain any obvious indications of a population 
older than the 1$-$2~Gyr old RC.  Although the models for the RGB suggest 
that there could be a globular cluster age old metal poor population.  
There is no obvious HB, especially no obvious BHB.  However, as is seen in 
Leo~I (Mateo {\it et al.} 1994; Gallart {\it et al.} in preparation), a 
lack of a HB is necessary but not sufficient proof that there is no old 
stellar component.  The Leo~I HST data shows distinct old MSTOs, but no 
corresponding HB.

\section{Conclusions}

Our data indicate that the majority of stars in the Leo~A dwarf irregular 
galaxy have ages of less than a few Gyr.  Is this reasonable? One way to 
test this conclusion is to see if other galaxies show independent evidence 
for star formation in the comparatively recent past.  For example, if our 
model is correct then we might expect to see a considerable range in mean 
stellar population ages among other extreme dwarf galaxies in the Local 
Group.  Since Leo~A is currently dim ($M_V = -$11.5 for our adopted 
distance; see Mateo 1998), we should be comparing it with the smallest 
dwarfs, which are mainly dwarf spheroidal (dSph) systems.

The stellar populations of Local Group dSph galaxies are being explored in 
detail through a combination of ground- and space-based observations (see 
reviews by Gallagher \& Wyse 1994, Da~Costa 1994, 1997, Mateo 1998).  These 
investigations reveal that several of the Galactic dSph contain large 
complements of intermediate age stars, e.g., Carina 
(Smecker-Hane {\it et al.}  
1994) and Leo I (Lee {\it et al.}  1993, Caputo, Castellani, \& Degl'Innocenti 
1996), while other dSph, such as the Ursa Minor system, are mainly composed 
of ancient, globular-cluster stellar populations.  Leo I has an especially 
prominent intermediate age stellar population component that was probably 
formed a few Gyr ago.  The dSph therefore display the kind of range in 
stellar ages that one might expect if major star-forming episodes have 
occurred sporadically in small galaxies during the past 10~Gyr or so.  That 
we are seeing Leo~A at a time when most of the stars have formed in the 
past few Gyr is evidently not a special case, but instead seems to be a 
relatively common phenomenon.

There remains a distinct possibility that Leo~A is a a purely young system.  
The ambiguities in interpreting the RGB mean that we still need to find 
unequivocal proof of an old population (e.g., RR~Lyr variable stars, or old 
MSTOs).  The new closer distance for Leo~A makes it possible to observe 
stars as faint as M$_V$=+4, equivalent to globular cluster age MSTOs.  
Stellar spectroscopy is also possible of RGB stars to try and look for 
evidence of metallicity evolution.  In any case, Leo~A is an excellent 
candidate for further study of star formation in a very low metallicity 
environment.  Even if it is does not contain solely a young population, it 
is still the nearest by example of what a FBG may look like.

\acknowledgements
Support for this work was provided by NASA through grant GO-5915 from the 
Space Telescope Science Institute, which is operated by AURA, Inc., under 
NASA contract NAS 3-26555.  Partial support from a NASA LTSARP grant no.  
NAGW-3189 is gratefully acknowledged.  We thank the referee for very useful 
comments which improved our presentation.  RD-P would like to thank the 
Graduate School of the University of Minnesota for a Dissertation 
Fellowship.  JSG expresses his appreciation for partial travel support from 
ST-ECF and ESO which made visits to Garching possible.  ET thanks Ralph 
Bohlin for research support in the early phases of this project, and 
Michael Rosa, Chris Burrows, Carme Gallart, Alfred Gautschy, Fred Lo and 
Cesare Chiosi for useful discussions.  This research has made use of the 
NASA/IPAC Extragalactic Database (NED) which is operated by the Jet 
Propulsion Laboratory, California Institute of Technology, under contract 
with NASA.
\vskip2cm

\appendix
\section{Variable Stars}

The instability strip, where stars become unstable and thus pulsate and 
vary in brightness, appears to go through the entire Hertzsprung-Russell 
diagram, (see Gautschy \& Saio 1995, Figure~1) and the basic properties are 
in principle well understood (e.g., Gautschy \& Saio 1995, 1996, and 
references therein).  The instability strip for the classical Cepheids 
($\delta$-Cepheids) continues down in mass to include RR~Lyrae, 
BL~Herculis, W~Virginis, RV~Tauri and even Mira variable stars (e.g., in 47 
Tuc).  The bluer and hotter Anomalous Cepheid (AC), are also in this 
instability strip at higher mass.

The $\delta$-Cepheids are $3 - 15 \Msun$ stars, whose evolutionary tracks 
take them through the HRD instability strip.  They vary uniformly and 
reliably obeying a tight well calibrated Period-Luminosity (PL) relation, 
which allows them to be one of the most important {\it primary} distance 
indicators (e.g., Sandage \& Tammann 1968).  At the very low mass end are 
the RR~Lyrae variable stars, which also have a well calibrated PL relation 
used for accurate distance determinations.  They are typically HB stars.  
In between RR~Lyrae and $\delta$-Cepheids is a rather poorly studied 
``gap'' (Hoffmeister, Richter \& Wenzel 1982), through which the 
instability strip passes.  Here lie the BL~Her and AC variables.

The literature contains very few studies of these ``non-classical'' (e.g., 
W~Vir, AC) variable stars (e.g., Nemec, Nemec \& Lutz 1994), certainly not 
in the studies of stellar populations outside our Galaxy halo or the 
Magellanic Clouds.  Early definitions of how to detect them is ``they will 
be 2.5 mags fainter than the classical Cepheids'' (Baade \& Swope 1963), 
and in general most prescriptions for identifying them continue to rest 
upon previous detections of $\delta$-Cephei or RR~Lyr variables, or some 
inherent knowledge of the properties of the region being searched (e.g., 
Wallerstein \& Cox 1984).  These non-classical Cepheids have often been 
called ``Population II Cepheids'' (to distinguish from $\delta$-Cepheids, 
or ``Population I Cepheids''), but this has long been known to be 
frequently an inaccurate wording (e.g., Woolley 1966; Richter 1967).

W~Vir and RV~Tau variable stars are thought to be post-HB stars that pass 
through the instability strip on their way to the AGB as they exhaust the 
helium in their cores (Gautschy \& Saio 1996; Sandage {\it et al.} 1994).  
Since the evolutionary time scale of the core helium exhaustion phase is 
much faster than the core helium burning phase, the number of these stars 
is typically much smaller than the RR~Lyr stars.  However this ratio might 
change in the case of a short burst of star formation.  It is also 
completely unknown if it is possible that young helium burning BL stars 
could cross into these same unstable region of parameter space at very low 
metallicities.  Young, extremely metal poor stars exist in a very under 
studied region of parameter space.  W~Vir stars are typically bluer than 
$\delta$-Cepheids (Fernie 1964; Dickens \& Carey 1967; Eggen 1961), and 
even have their own Period-Luminosity relation (Eggen 1961):

$$ M_V = -0.13 - 1.90 \log P $$

This relation has a different slope from that of classical Cepheids, and 
this means they can be between 1 and 2.5 magnitudes brighter than their 
classical cousins, depending upon their pulsation period.

ACs have shorter periods and are more massive than W~Vir stars and are 
known to have two possible separate evolutionary paths, neither of which 
comes from a post-HB phase (e.g., Bono {\it et al.} 1997; Smith \& Styker 
1986).  They can be either single stars, thus much younger than the RR~Lyr 
population, at 1$-$4~Gyr old, or they can be binary stars where mass 
transfer has taken place.  In this last case, for example a binary of two 
stars of 0.9$\Msun$ can be as old as $\sim$10~Gyr (Renzini, Sweigart \& 
Mengel 1977).  They have their own PL relation, with a metallicity 
dependence (Nemec {\it et al.} 1994), which is, for the fundamental 
pulsation mode:

$$ M_V = -0.10 - 3.13 (\pm 0.28) \log P + 0.32\rm[Fe/H]$$

The SMC is known to contain a number of ACs (Graham 1975), whereas none (or 
few) have been detected in the LMC (e.g., Graham 1975, 1977, 1985).  As the 
LMC and SMC are thought to have similar, not to say inextricably linked, 
SFHs, this has been suggested to be a metallicity effect (Smith \& Stryker 
1986).  Thus, one might expect an unusually large number of ACs in Leo~A 
which is an extremely metal poor galaxy.  Thus, we cannot take the 
identification of ``Pop II'' Cepheids as conclusive proof of globular 
cluster age stars because there are typically old and young evolutionary 
paths to these variable stars.

Classifying variable stars is a very complex and difficult especially with 
only a single filter, but since we can show that these variables in Leo~A 
follow the W~Vir PL relation, and if this can proven to be the case (with 
better sampling of the light curves), then we leave it as a problem for the 
theorists to explain why we observe that stars of these masses entering the 
instability strip.

The uncertainties in the different populations of variable stars which we 
have found calls into question the efficacy of using small numbers of 
Cepheid variable stars in either small galaxies with a peculiar SFH, and/or 
in galaxies with extremely low metallicity ($\sim$2\% solar) without a lot 
of care and attention to properly sampling light curves in more than one 
colour.  This conclusion also highlights the dangers of not considering the 
consequences of ``peculiar'' SFHs on the stellar population of a galaxy as 
a whole.  If it is assumed that a galaxy has been forming stars 
continuously for a number of Gyrs when in fact there have been several 
short epoches of star formation then several properties and their 
implications maybe be misjudged.  Distance indicators, such as Cepheid 
variable stars, have been designed and proven for large galaxies which 
typically have had fairly constant SFR over the past few Gyr.  If they are 
applied in other circumstances without due care and attention then problems 
may occur.

\clearpage
{\bf Tables}

Table~1: HST Photometric Calibration

Table~2: Properties of Leo~A

Table~3: Distance Estimates to Leo~A

Table~4: WFPC2 Observations of Cepheids in Leo~A

\clearpage


\begin{figure}
\caption{ WFPC2 footprint superimposed on 15' square piece of
the POSS where Leo~A is faintly visible as a smudge in the centre.  The 
WFPC2 pointing clearly covers most of the central star forming region of 
the galaxy.  The HI contours of the total HI map of Young \& Lo~(1997) 
covers this whole area.
\label{POSS} }
\end{figure}
\vskip0.3cm

\begin{figure}
\caption{ Mosaic of the WFPC2 F555W (`V') images of the
Leo~A dwarf irregular galaxy.  The galaxy is highly resolved and 
transparent.  A number of background galaxies can be seen {\it through} 
Leo~A.  Stellar crowding is clearly not a factor in the analysis of these 
images.
\label{MOS}}
\end{figure}
\vskip0.3cm

\begin{figure}
\caption{ 
\vskip0.2cm
Observed CMDs for Leo~A, plotted by chip for each of the 4 separate WFPC2 
CCDs.  We plot V magnitude vs.  B-V colour.  These come from 
transformations of the F555W and F439W filters.  The numbers of stars 
measured on each chip are printed within each CMD.
\label{CHIP1}} 
\end{figure}

\begin{figure}
\caption{ 
\vskip0.2cm
The same as figure~\ref{CHIP1}, except we plot I magnitude versus V-I 
colour.  These are transformations from F555W and F814W WFPC2 filters.  The 
numbers of stars measured on each chip are printed in each CMD.
\label{CHIP2}} 
\end{figure}
\vskip0.3cm

\begin{figure}
\caption{ 
\vskip0.2cm
The combined CMD of all four WFPC2 chips, in V, B-V.  The number of stars 
making up this CMD is 2636 stars matched in both filters.
\label{CMD1}} 
\end{figure}
\vskip0.3cm

\begin{figure}
\caption{ 
\vskip0.2cm
The combined CMD of all four WFPC2 chips, in I, V-I.The number of stars 
making up this CMD is 7295 stars matched in both filters.
\label{CMD2}} 
\end{figure}
\vskip0.3cm

\begin{figure}
\caption{ These plots show the DoPHOT estimated (thus conservative)
photometric errors for all the stellar
magnitudes in each 
of the three filters used.
\label{ERR}} 
\end{figure}
\vskip0.3cm

\begin{figure}
\caption{This is the MDM Hiltner 2.5m telescope 
calibrated CMD in I, V$-$I used to check the calibration zero points of the 
HST data.  The plot scale is the same as the HST CMD (figure~\ref{CMD2}),
but the 
MDM CCD covers more area on the sky, and thus
of Leo~A than HST. This means that the brighter sparsely distributed
populations are better represented in this CMD (e.g.
the TRGB). 
\label{MDMCMD}} 
\end{figure}
\vskip0.3cm

\begin{figure}
\caption{Here we show the MDM vs. HST photometric comparison for 15
stars that we found to be relatively isolated in the HST F814W chip~3 
image.  Chip~3 is the most crowded of the chips, and was chosen
because it gives us the largest sample of comparison stars.  
In the upper plot we make 
the comparison for V and below this I and in this bottom plot we show how 
the colour of the stars appears to vary between the two calibrations.  In 
the upper two plots the cross in a square symbol represents the reddest 
stars in the sample, that have V$-$I$>$0.8.  In the bottom plot this symbol 
represents the brighter stars in the sample, where I$<$21.
\label{MDMHST}} 
\end{figure}
\vskip0.3cm

\begin{figure}
\caption{Here we show the spatial distribution in stellar density
of differently
selected stellar populations in Leo~A, as seen in the HST images of Leo~A. 
\vskip0.2cm
The top panel shows the density distribution of the MS+BL (young, blue stars).
\vskip0.2cm
The middle panel shows the RC stars distribution.
\vskip0.2cm
The bottom panel shows the distribution of the RGB stars above the RC 
\vskip0.2cm
The stellar density contours are labeled in the maps in units of number
of stars per kpc square area.
\label{RDP1}} 
\end{figure}
\vskip0.3cm

\begin{figure}
\caption{ Here we show the spatial density distribution of the red and blue
stellar populations seen in
the MDM images of Leo~A.
\vskip0.2cm
The top panel shows the density distribution of the MS+BL (young, blue stars). 
\vskip0.2cm
The bottom panel shows the distribution of the RGB stars. 
\vskip0.2cm
The stellar density contours are labeled in the maps in units of number
of stars per kpc square area.
\label{RDP2}} 
\end{figure}
\vskip0.3cm

\begin{figure}
\caption{ We have over-plotted 
RGB Sequences from Da~Costa \& Armandroff 1990 for 5 globular clusters 
(M15, NGC6397, NGC6752, NGC1851, 47~Tuc) on our Leo~A data.  These cluster 
have, respectively, [Fe/H] = (-2.17, -1.91, -1.54, -1.29, -0.71).  They are 
plotted from right to left, with the lower metallicities the bluest.  We 
have done this for two different distance moduli to show the uncertainty as 
well as the implications of distance errors on the interpretation of these 
observations.
\label{RGB}} 
\end{figure}
\vskip0.3cm

\begin{figure}
\caption{ Here we show 3 images through Leo~A, 
which are taken from the WF3 chip.  They are from top to bottom, in F814W, 
F555W and F439W filters.  Each image is 16'' square (40~pc at distance of 
Leo~A), and their limiting magnitudes are roughly similar (I=24.8, V=25.8 
and B=25.3).
\label{EXT}} 
\end{figure}
\vskip0.3cm

\begin{figure}
\caption{ 
\vskip0.2cm
a.  Our optimum model for the young stellar pop ($<$1.5~Gyr old), Z=0.0004 
metallicity and assuming constant star formation (in red), over plotted on 
the data (black)
\vskip0.2cm
b.  Our optimum model for the young stellar pop ($<$1.5~Gyr old), Z=0.0004 
metallicity and assuming a burst of star formation ended 900~Myr ago, and 
began at least 1.5~Gyr ago, and another epoch of star formation began 
500Myr ago and continues to the present in red.  The data is in black.
\vskip0.2cm
The main difference between these two plots is seen at the faint end of the 
CMD - where the burst model produces more stars along the faint limit of 
the observations.  The difference is very slight to the eye, but it 
significantly improves various aspects of the model/data comparison.  Which 
highlights the importance of numerical comparisons between models and data.
\label{YG}} 
\end{figure}
\vskip0.3cm

\clearpage

\begin{figure}
\caption{Here we show the SFR versus time derived using the Dohm-Palmer
method from the luminosity function for stars identified as being on the 
core-HeB phase.  The bins are 100~Myr wide.  The data have been dereddened 
and corrected for incompleteness.  The assumed IMF has a Salpeter slope.  
Error bars refer only to statistical uncertainties due to the number of 
stars in each bin.
\label{RDP3}} 
\end{figure}
\vskip0.3cm

\begin{figure}
\caption{
\vskip0.2cm
a.  Our optimum model of the RC ($\sim$1$-$2~Gyr old) in red superimposed 
upon the data (black points).
\vskip0.2cm
b.  Our optimum model of the RGB ($\sim$9$-$10Gyr) in red superimposed upon 
the data (black points).
\vskip0.2cm
Clearly these models cannot be made consistent with each other.
\label{INT}} 
\end{figure}
\vskip0.3cm

\begin{figure}
\caption{Our overall optimum model, where the SFH that this presumes is
shown in figure~\ref{SFH}.  The top plot shows the complete model, in the 
bottom panel the different age star formation epochs that contributed to 
this distribution are separated out.  The young population (0$-$500~Myr) is 
shown in red, the intermediate age ``burst'' of star formation 
(900$-$1500~Myr) is shown in green, and the possible underlying old 
population (9$-$10~Gyr) is shown in blue.
\label{MOD}}
\end{figure}
\vskip0.3cm

\begin{figure}
\caption{ Here we show one
possible self consistent SFH for Leo~A which matches most of the data very 
well.  We have not been able to estimate the effects of metallicity 
evolution because we lack a consistent set of models at metallicities lower 
than those used for our youngest stars.  As described in the text the 
errors are not statistical, they merely give an indication of how flexible 
the SFR at any time can be before significantly affecting the good match of 
this model to the data.
\label{SFH}} 
\end{figure}

\end{document}